\documentclass[prb,twocolumn,amsmath,amssymb,superscriptaddress, bibnotes]{revtex4-2}

\usepackage{graphicx}
\usepackage{dcolumn}
\usepackage{bm}
\usepackage{color}
\usepackage{braket} 

\begin{document}

\title{Breakdown of Kubo relation in Pt-Cu nanoparticle}

\author{Shunsaku~Kitagawa}
\email{kitagawa.shunsaku.8u@kyoto-u.ac.jp}

\author{Yudai~Kinoshita}
\author{Kenji~Ishida}

\affiliation{Department of Physics, Kyoto University, Kyoto 606-8502, Japan}

\author{Kouhei~Kusada}

\affiliation{Department of Chemistry, Kyoto University, Kyoto 606-8502, Japan}
\affiliation{The Hakubi Center for Advanced Research, Kyoto University, Kyoto 606-8501, Japan}

\author{Hiroshi~Kitagawa}

\affiliation{Department of Chemistry, Kyoto University, Kyoto 606-8502, Japan}

\date{\today}

\begin{abstract}
Nanoparticles were predicted to exhibit unique physical properties due to quantum size effects, but their identification remains difficult.
According to Kubo's theory, the gap size is inversely correlated with both the density of states at the Fermi energy and the number of atoms in the particle.
Previously, we confirmed that the particle size and magnetic field dependence of NMR anomaly temperature is consistent with the estimated ``Kubo'' gap.
Here, we investigated the density-of-states dependence in the Pt$_{1-x}$Cu$_{x}$ nanoparticles.
While an enhancement of nuclear spin-lattice relaxation rate $1/T_1$ at low temperatures was clearly observed for the Pt-rich nanoparticles, such behavior was abruptly suppressed in the Cu-rich nanoparticles.
Furthermore, the NMR anomaly temperature is nearly unchanged with varying the density of states.
Our findings indicate that the quantum size effect contains more profound physics than just the ones predicted by Kubo.
\end{abstract}

\maketitle

Nanoparticles, also known as quantum dots, are particles with diameters on the order of nanometers\cite{A.I.Ekimov_JETPL_1981,Rossetti1983}.
They have been studied\cite{R.Kubo_JPSJ_1962,L.P.Gorkov_JETP_1965,W.P.Halperin_RMP_1986} and utilized in a wide range of fields\cite{A.P.Alivisatos_science_1996,Seh2017,V.Harish_nano_2022,Antoine2023} because they exhibit properties that are different from those of bulk materials, such as an extremely large surface area and unique physical properties due to the quantum size effect (QSE).
In the early 1960s, Kubo theoretically predicted that the physical properties of metal nanoparticles differ from those of the bulk, depending on the number of atoms they contain\cite{R.Kubo_JPSJ_1962}.
Bulk metals have a continuous band dispersion, but as the number of atoms decreases, the energy dispersion becomes discontinuous and the energy gap opens in nanoparticles.
Because of the discrete energy levels, for instance, nanoparticles with even and odd numbers of electrons show different magnetic susceptibilities\cite{R.Kubo_JPSJ_1962,J.Sone_JPSJ_1977}.
According to Kubo's theory, the mean gap size $\delta_{\rm Kubo}$ was determined by the equation of $\delta_{\rm Kubo}~=~[ND(E_{\rm F})]^{-1}$.
Here, $N$ is the number of atoms in a nanoparticle, and $D(E_{\rm F})$ is the density of states at Fermi energy $E_{\rm F}$.

Even though there has been much numerous research on nanoparticles\cite{P.Yee_PRB_1975,T.Goto_JPSJ_1989,C.D.Makowka_PRB_1985,Bucher1989,vanderKlink2000,Marbella2015,T.Fujii_PRB_2022}, it is still challenging to identify the QSE.
This is because the surface effect, different from the bulk behavior, is similar to that expected by the QSE.
Because of ``a boring surface condition'', such as oxidation or degradation by coating materials, the surface areas of nanoparticles often become nonmetallic.

In previous reports, we investigated the electronic properties of Pt nanoparticles with $^{195}$Pt-nuclear magnetic resonance (NMR) measurements\cite{T.Okuno_PRB_2020,T.Okuno_JPSJ_2020}.
The NMR signals arising from the surface and the inner areas of the nanoparticles could be distinguished owing to the large $D(E_{\rm F})$ of Pt in our previous measurements.
We found that the nuclear spin-lattice relaxation rate $1/T_1$ of the nanoparticles deviates from the bulk behavior and increases below the characteristic temperature $T^{*}$.
These NMR anomalies were observed in both the surface and interior signals, indicating that this is not a surface effect but QSE.
Moreover, the systematic particle size and magnetic field dependencies strongly imply that $k_{\rm B}T^{*}$ would correspond to the ``Kubo'' discrete-energy gap.
According to Kubo's theoretical prediction, $\delta_{\rm Kubo}$ depends not only on the particle size but also on $D(E_{\rm F})$.
Therefore, to further clarify the QSE, the systematic $D(E_{\rm F})$ dependence of  $T^{*}$ is important.

\begin{figure}[!tb]
\includegraphics[width=8.5cm,clip]{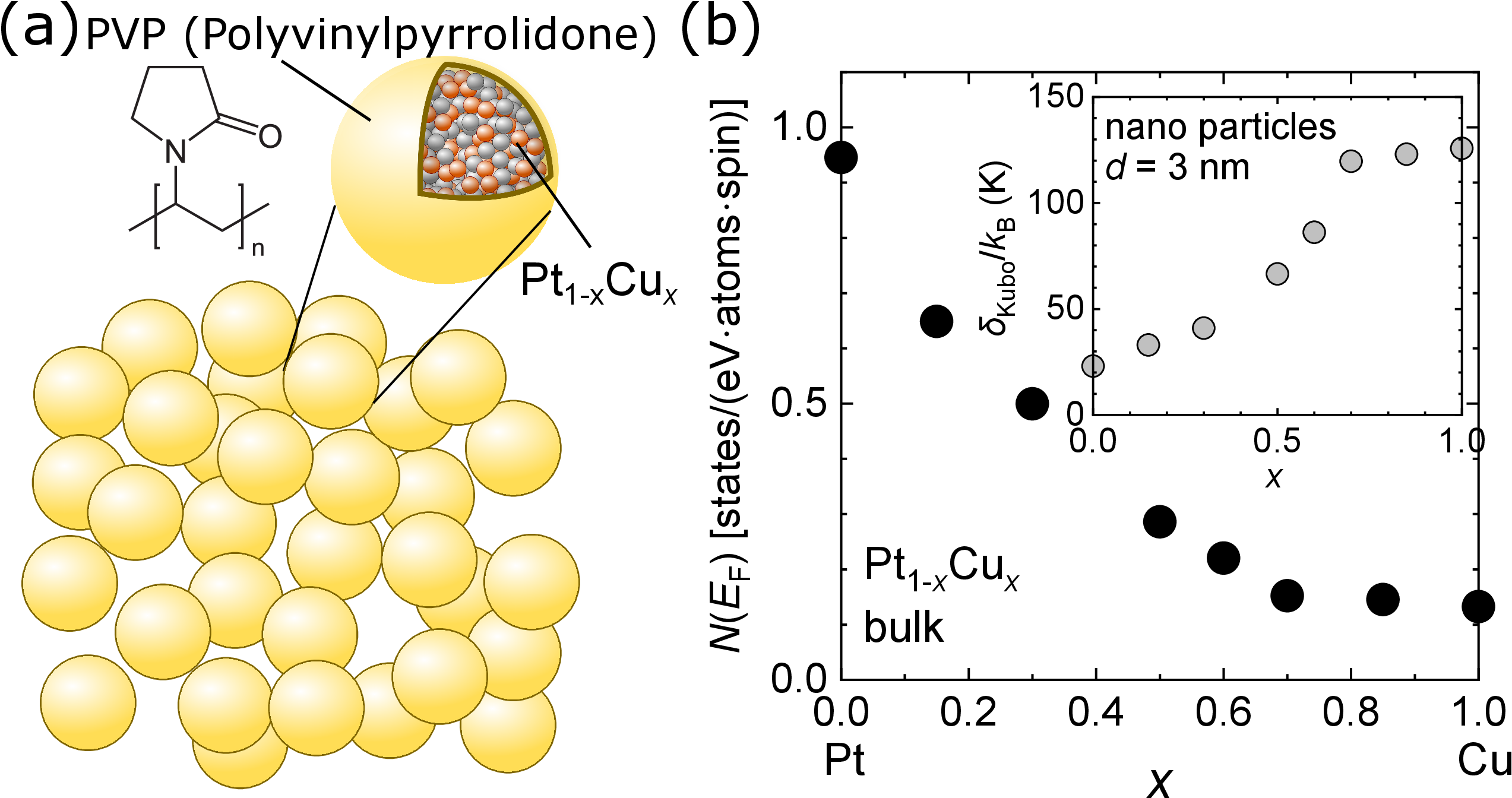}
\caption{
(a) Schematic image of Pt$_{1-x}$Cu$_{x}$ nanoparticles.
Nanoparticles are covered by polyvinylpyrrolidone to prevent oxidation and particle-to-particle contact. 
(b) Cu content $x$ dependence of the density of states at Fermi energy $D(E_{\rm F})$ in bulk Pt$_{1-x}$Cu$_{x}$ alloy.\cite{PhysRevB.40.12079}
(Inset) $x$ dependence of theoretically estimated energy gap $\delta_{\rm Kubo}/k_{\rm B}$ in nanoparticles of 3 nm diameter.
$\delta_{\rm Kubo}$ is calculated by the equation of $\delta_{\rm Kubo}~=~[ND(E_{\rm F})]^{-1}$.
Here, $N$ is the number of atoms in a nanoparticle.
}
\label{Fig.1}
\end{figure}

\begin{figure*}[!tb]
\includegraphics[width=\linewidth,clip]{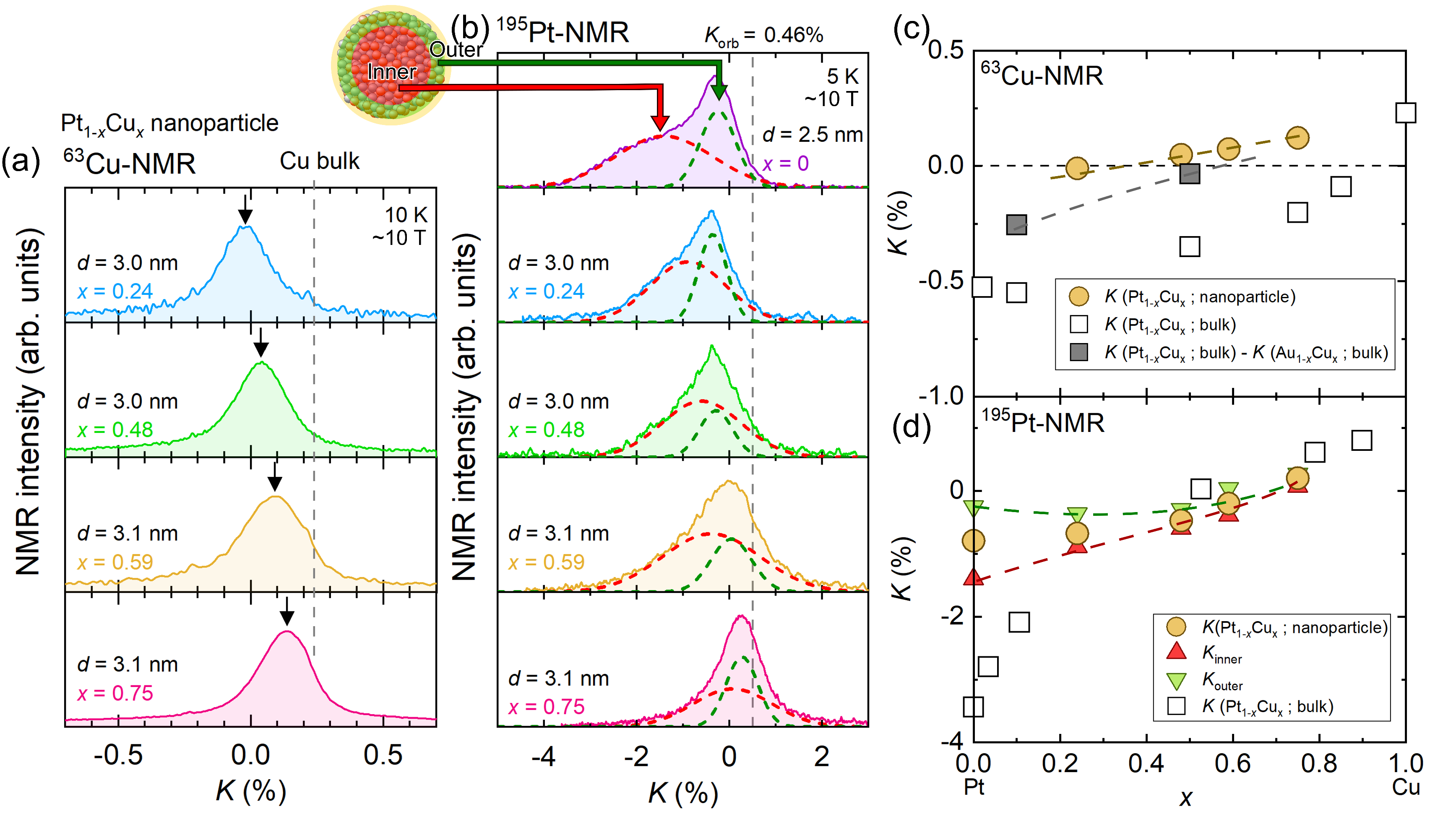}
\caption{
$^{63}$Cu- (a) and $^{195}$Pt- (b) NMR spectrum of Pt$_{1-x}$Cu$_{x}$ nanoparticles for various $x$.
The measurement temperature and magnetic field are indicated in the panel.
For $^{63}$Cu-NMR, the arrows indicate the peak position of the spectrum.
The dashed line indicates the Knight shift value of Cu bulk.
For $^{195}$Pt-NMR, the two-peak structure originating from the inner and the outer atoms was observed, indicated by the dashed curves.
The dashed line indicates the orbital part of Knight shift $K_{\rm orb}$~=~0.46\%.
The $x$ dependence of the peak position of Knight shift $K$ for $^{63}$Cu- (c) and $^{195}$Pt- (d) NMR.
For comparison, we also plot $K$ in bulk.
For $^{63}$Cu-NMR in bulk, we subtract the contribution of quadrupole interaction using the value of Au$_{1-x}$Cu$_{x}$ alloy.
The broken curves indicate the guide for eyes.
}
\label{Fig.2}
\end{figure*}

Here, we performed $^{63}$Cu- and $^{195}$Pt-NMR measurements on Pt$_{1-x}$Cu$_{x}$ alloy nanoparticles with a similar particle size (3~nm).
As shown in Fig.~\ref{Fig.1}(b), by mixing Pt and Cu, the density of states can be varied continuously\cite{PhysRevB.40.12079} and $\delta_{\rm Kubo}$ is expected to change as well.
We found that NMR results are almost identical in $^{63}$Cu- and $^{195}$Pt-NMR, indicating that those reflect QSE.
For Pt-rich nanoparticles, an increase in $1/T_1$ below $T^{*}$ was clearly observed, as in the case for Pt nanoparticles, while such behavior was abruptly suppressed at around $x = 0.5$ and was not observed for Cu-rich nanoparticles.
Furthermore, $T^{*}$ does not change so much with varying the density of states, suggesting that $T^{*}$ is not simply proportional to $\delta_{\rm Kubo}$.

\begin{figure}[!tb]
\includegraphics[width=\linewidth,clip]{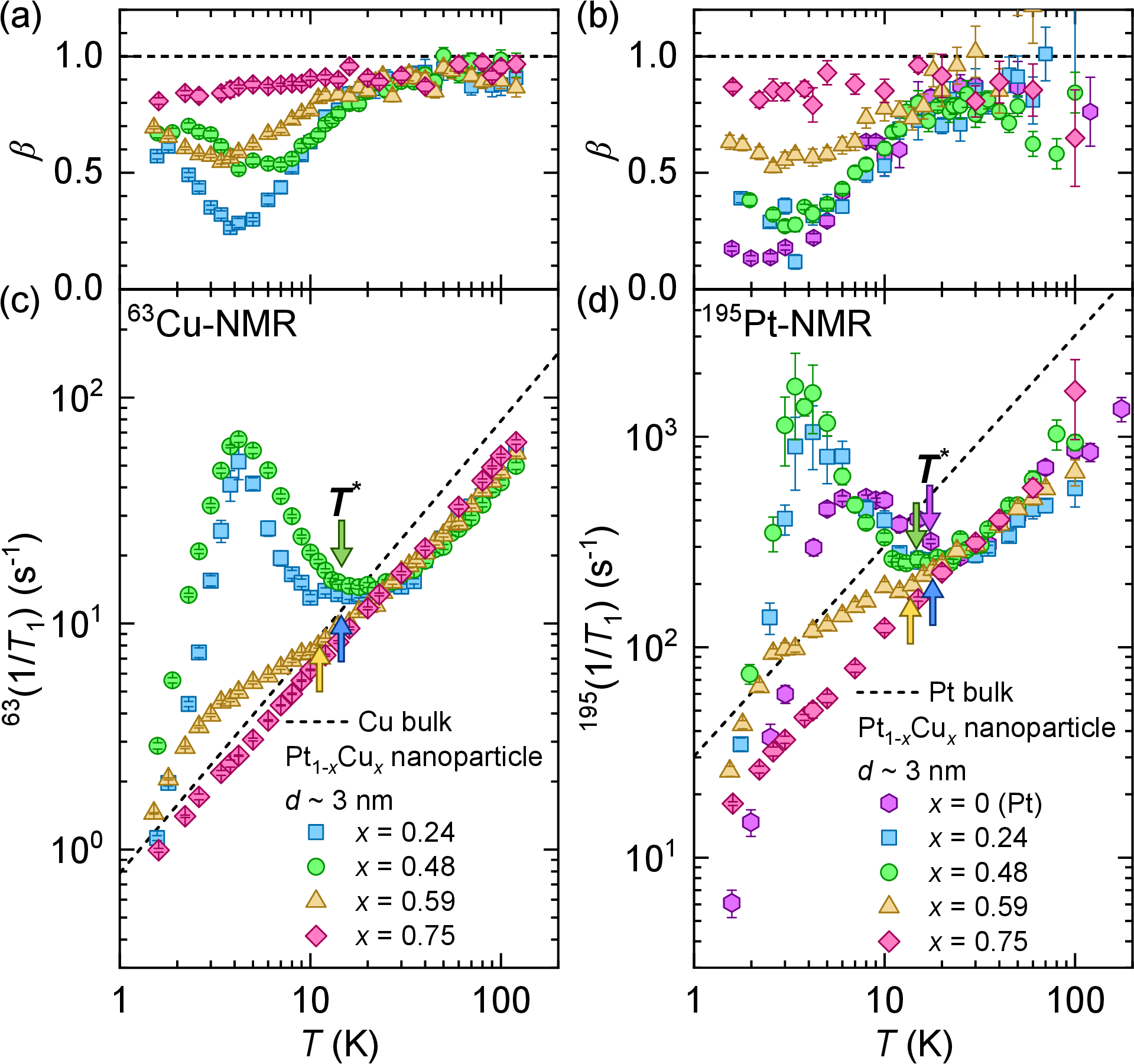}
\caption{
Temperature dependence of stretched exponent $\beta$ for $^{63}$Cu-(a), and $^{195}$Pt-NMR (b).
The dashed lines indicate unity (the $T_1$ can be evaluated through the single exponential fitting).
Temperature dependence of $1/T_1$ at various $x$ for $^{63}$Cu- (c) and $^{195}$Pt-NMR (d).
The dashed lines indicate bulk value.
The arrows indicate $T^*$, where $1/T_1$ starts to increase.
}
\label{Fig.3}
\end{figure}

In a typical synthesis of Pt$_{1-x}$Cu$_{x}$ nanoparticles ($x = 0.48$) [Fig.~\ref{Fig.1}(a)], poly(N-vinyl-2-pyrrolidone) (PVP K-30, 1110 mg, Wako) was dissolved in triethylene glycol (TEG, 250 ml, Wako).
NaOH (6.0 mmol, Wako) dissolved in 2 ml of H$_2$O was added to the TEG solution and the solution was heated to 225 $^{\rm o}$C in the air with magnetic stirring.
Meanwhile, K$_2$[PtCl$_4$] (1.0 mmol, Aldrich) and CuCl$_2$·2H$_2$O (1.0 mmol, Wako) were dissolved in deionized water (20 ml).
The aqueous mixture solution was then slowly added to the heated TEG solution.
The solution was maintained at 220 $^{\rm o}$C while adding the solution.
After cooling to room temperature, the prepared nanoparticles were separated by centrifuging.
Other Pt$_{1-x}$Cu$_{x}$ ($x = 0.24, 0.59$, and 0.75) nanoparticles were prepared by controlling the molar ratio of Pt$^{2+}$ and Cu$^{2+}$ ions.
The details of the synthesis conditions for Pt$_{1-x}$Cu$_{x}$ nanoparticles are summarized in Tab~\ref{tab.1}.
\begin{table}[!tb]
    \centering
\caption{Reaction Condition for the syntheses of Pt$_{1-x}$Cu$_{x}$ nanoparticles.}
\label{tab:my_label}
    \begin{tabular}{cccccc} \hline
         Sample&  \begin{tabular}{c} K$_2$[PtCl$_4$] \\(mmol)  \end{tabular}&  \begin{tabular}{c} CuCl$_2$·2H$_2$O \\(mmol)  \end{tabular}&  \begin{tabular}{c} TEG \\(ml)  \end{tabular}&  \begin{tabular}{c} PVP \\(mmol)  \end{tabular} & \begin{tabular}{c} NaOH \\(mmol)  \end{tabular}
\\ \hline 
         Pt$_{0.76}$Cu$_{0.24}$&  1.5&  0.5&  250&  10.0& 6.0
\\ 
         Pt$_{0.52}$Cu$_{0.48}$&  1.0&  1.0&  250&  10.0& 6.0
\\ 
         Pt$_{0.41}$Cu$_{0.59}$&  0.8&  1.2&  250&  10.0& 6.0
\\ 
         Pt$_{0.25}$Cu$_{0.75}$&  0.5&  1.5&  250&  10.0& 15.0
\\ \hline
    \end{tabular}
\label{tab.1}
\end{table}
All the samples have a face-centered cubic crystal structure, as determined by an X-ray diffraction measurement.
The average diameter of the nanoparticles, as described in Fig.~\ref{Fig.2}, was measured from transmission electron microscopy images.
A nanoparticle of 3~nm in diameter contains $\sim$ 600 atoms.
The Cu content $x$ in nanoparticles was determined by energy dispersive X-ray spectroscopy (EDS).
EDS indicates the homogeneous mixture of Pt and Cu atoms in each nanoparticle.
The NMR measurements were performed using a standard spin-echo method.
The $^{63}$Cu (nuclear spin $I~=~3/2$, nuclear gyromagnetic ratio $\gamma/2\pi~=~11.285$~MHz/T, and natural abundance 69.1\%)- and $^{195}$Pt (nuclear spin $I~=~1/2$, nuclear gyromagnetic ratio $\gamma/2\pi~=~9.153$~MHz/T, and natural abundance 33.8\%)-NMR spectra were obtained as a function of the magnetic field at a fixed frequency ($f_0$ = 111~MHz).
The magnetic field was calibrated using $^{63}$Cu and $^{65}$Cu ($^{65}\gamma_n/2\pi = 12.089$~MHz/T, and natural abundance 30.9\%) NMR signals from the NMR coil and was then converted to a Knight shift as $K (\%) = (f_0/\gamma - H)/H \times 100$.
$T_1$ was measured using the saturation-recovery method.
The single-component $T_1$ was evaluated through the exponential fitting in the high-temperature range where the Korringa relation holds.
At low temperatures, the recovery of the nuclear magnetization shows a multi-exponential behavior, and thus the fitting was performed by the stretched exponential function described as,
\begin{align}
    M(t)  = M(\infty)\left[ 1 - \exp\left\{ - \left( \frac{t}{T_1}\right)^\beta \right\}\right].
\end{align}
Here, $M(t)$ is the time-dependent nuclear magnetization and $\beta$ is an exponent related to the inhomogeneity of the relaxation component.
Note that $\beta = 1$ means the single $T_1$ relaxation and $\beta$ becomes smaller than 1 when relaxation becomes multicomponent.
The temperature dependence of stretched exponent $\beta$ is shown in Figs.~\ref{Fig.3}(a) and \ref{Fig.3}(b).

The systematic change in the NMR spectrum with varying $x$ is clearly observed as shown in Figs.~\ref{Fig.2}(a) and \ref{Fig.2}(b).
Since the NMR Knight shift is proportional to local magnetic susceptibility at the observed nuclear sites, the NMR peak and shape reflect the average magnetic susceptibility and the distribution of magnetic susceptibility, respectively.
For $^{63}$Cu-NMR, a single-peak structure was observed as shown in Fig.~\ref{Fig.2}(a), suggesting the relatively uniform distribution of magnetic susceptibility due to the small bulk susceptibility.
A peak position shifts to a higher $K$ side with increasing $x$ and approaches the bulk-$K$ position\cite{J.Itoh_1964}.
On the other hand, the $^{195}$Pt-NMR spectrum in the Pt-rich region was broad [Fig.~\ref{Fig.2}(b)].
This is because the inner (bulk-like) signal has a larger magnetic susceptibility while the outer (surface) signal has a smaller magnetic susceptibility.
Because the $D(E_{\rm F})$ of Pt is much larger than that of Cu due to the presence of $d$ electrons, the distribution of $K$ in the nanoparticle is also larger in Pt-NMR than in Cu-NMR.
A double-Gaussian function was used for the estimation of the average magnetic susceptibility of the inner and outer parts.
The difference between the inner and outer magnetic susceptibilities decreases with increasing $x$, and almost a single NMR peak was observed in the Cu-rich $x=0.75$ sample.
Note that the position and shape of the $^{63}$Cu- and $^{195}$Pt-NMR spectrum are almost independent of temperature.
This means that the effect related to the odd/even number of electrons could not be observed in the present measurement.
We compare the $x$ dependence of $K$ of $^{63}$Cu- and $^{195}$Pt-NMR in nanoparticles with that in bulk\cite{J.Itoh_1964}, as shown in Figs.~\ref{Fig.2}(c) and \ref{Fig.2}(d), respectively.
For $^{63}$Cu-NMR in the bulk, the contribution of quadrupole interaction for $K$ was subtracted using the value of Au$_{1-x}$Cu$_{x}$ alloy\cite{J.Itoh_1964}.
The $x$ dependence of $K$ in nanoparticles is consistent with that in the bulk, indicating the $x$ dependence of $D(E_{\rm F})$ is similar to each other.

\begin{figure}[!tb]
\includegraphics[width=0.85\linewidth,clip]{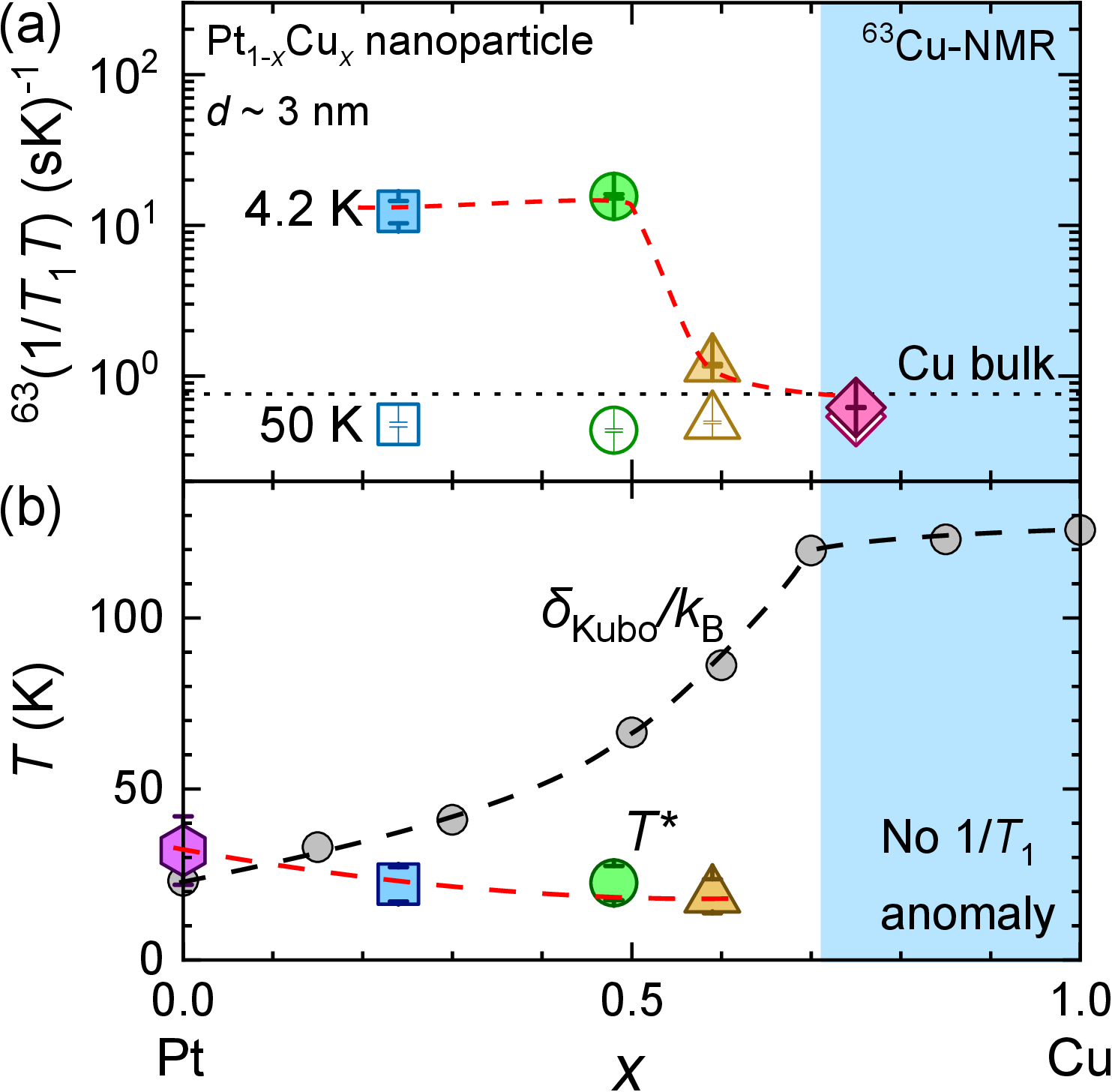}
\caption{
$x$ dependence of $1/T_1T$ for $^{63}$Cu-NMR at 4.2~K and 50~K (a) and $T^*$ together with $\delta_{\rm Kubo}/k_{\rm B}$ (b) in Pt$_{1-x}$Cu$_{x}$ nanoparticles.
The dotted line indicates the bulk value.
The dashed curves indicate the guide for eyes.
No enhancement of $1/T_1T$ was observed in the high $x$ region indicated by the colored box.
}
\label{Fig.4}
\end{figure}

Figures~\ref{Fig.3}(a) and \ref{Fig.3}(b) show the temperature dependence of $^{63(195)}(1/T_1)$ at various $x$ for $^{63}$Cu($^{195}$Pt)-NMR, respectively.
Both $1/T_1$'s show similar temperature dependence, reflecting the intrinsic nature of the nanoparticle as a whole.
In the high-temperature regions, $^{63,195}(1/T_1)$ is proportional to temperature in all samples, indicating metallic behavior in nanoparticles.
As $^{63,195}(1/T_1)$ was measured at a lower $K$ position than bulk, the value of $^{63,195}(1/T_1)$ for the nanoparticles is smaller than that for the bulk.
In the Pt-rich samples ($x < 0.48$), $^{63,195}(1/T_1)$ deviates from the bulk behavior and increases below $T^{*}$, similar to the Pt nanoparticles.
On the other hand, such an enhancement of $^{63,195}(1/T_1)$ is suppressed with increasing $x$, and $1/T_1$ is almost proportional to temperature at $x = 0.75$, suggesting that the $^{63,195}(1/T_1)$ anomaly at low temperatures is caused by the contribution of Pt.
To clarify the variation of the low-temperature anomaly with Cu content, $x$-dependence of $^{63}(1/T_1T)$ at 4.2~K and 50~K is plotted in Fig.~\ref{Fig.4}(a).
The $x$ dependence of $T^{*}$ and the estimated Kubo temperature $\delta_{\rm Kubo}/k_{\rm B}$ from the bulk  $D(E_{\rm F})$ are also plotted in Fig~\ref{Fig.4}(b). 
The peak height of $^{63}(1/T_1T)$ is found to be nearly constant below $x = 0.5$ and is immediately suppressed for $x > 0.5$. 

The most striking point is that $T^{*}$ is almost independent of $x$, as shown in Fig.~\ref{Fig.4}(b).
From a theoretical point of view, $\delta_{\rm Kubo}$ is proportional to the inverse of $D(E_{\rm F})$, and thus, $\delta_{\rm Kubo}$ should increase with decreasing $D(E_{\rm F})$.
In Pt$_{1-x}$Cu$_{x}$ nanoparticles, it was experimentally confirmed that $D(E_{\rm F})$ decreases with increasing $x$, as shown in Fig.~\ref{Fig.2}(d).
Therefore, it is naively expected that $T^{*}$, which is related to $\delta_{\rm Kubo}$, increases with increasing $x$.
However, it was found that the experimentally estimated $T^{*}$ does not follow our expectation.

The above experimental results suggest that the low-temperature NMR anomaly in the nanoparticles is not a simple ``Kubo'' effect.
Here, we summarize the characteristics of the NMR anomalies.
In the Pt$_{1-x}$Cu$_{x}$ nanoparticles, the almost identical temperature dependence of $^{63,195}(1/T_1)$ implies that the relaxation process at both atomic sites is determined with the same conduction electrons.
According to the previous reports\cite{T.Okuno_PRB_2020,T.Okuno_JPSJ_2020}, the particle size and magnetic field dependencies strongly suggest that $T^{*}$ would correspond to $\delta_{\rm Kubo}$.
On the other hand, the low-temperature anomaly disappears, and $T^{*}$ is independent of $D(E_{\rm F})$ in the Cu-rich Pt$_{1-x}$Cu$_{x}$ nanoparticles.
The systematic change of Knight shift is consistent with that of $D(E_{\rm F})$, but the abrupt suppression of the low-temperature $1/T_1T$ anomaly suggests the existence of an $x$ concentration threshold for the observation of such anomaly.
Since $D(E_{\rm F})$ of Pt is much larger than that of Cu, we consider that the Kubo effect in the Pt$_{1-x}$Cu$_{x}$ nanoparticles is governed by the dominance of the Pt-5$d$ electrons.

Another important factor is the difference in $D(E_{\rm F})$ between the inner and the outer atoms.
As shown in Fig.~\ref{Fig.2}(b), the $^{195}$Pt-NMR spectrum in the Pt-rich Pt$_{1-x}$Cu$_{x}$ nanoparticles exhibits two distinguishable peaks originating from the inner and the outer atoms.
On the other hand, for $x \ge 0.5$ where the low-temperature anomaly is suppressed, the peak positions of two spectra are close.
This indicates that the distribution of $D(E_{\rm F})$ in the Pt$_{1-x}$Cu$_{x}$ nanoparticles becomes small for $x \ge 0.5$. 
While the specific mechanism remains unclear, it is pointed out that the significant difference in $D(E_{\rm F})$ between the inner and the outer atoms plays a crucial role in the low-temperature anomaly.
In any cases, it seems that the present observation cannot be interpreted with the widely-believed Kubo theory but suggests the necessity of the modification of the theory to explain the observation.

In conclusion, we performed $^{63}$Cu- and $^{195}$Pt-NMR in the Pt$_{1-x}$Cu$_{x}$ nanoparticles.
While an enhancement of nuclear spin-lattice relaxation rate $1/T_1$ was clearly observed for the Pt-rich nanoparticles, such behavior was suppressed in the Cu-rich nanoparticles, indicating the enhancement is related to the Pt $d$ electrons.
Furthermore, $T^{*}$ does not change so much with varying $x$, i.e. the density of states.
Our observations indicate that the understanding of QSE requires new mechanisms other than the widely believed ``Kubo'' effect.

\section*{acknowledgments}
The authors would like to thank Y. Maeno and S. Yonezawa for their valuable discussions.
This work was partially supported by the Kyoto University LTM Center and Grants-in-Aid for Scientific Research (KAKENHI) (Grants No. JP20KK0061, No. JP20H00130, No. JP20H05623, No. JP21K18600, No. JP22H04933, No. JP22H01168, and No. JP23H01124).

\end{document}